\documentclass[12pt]{article}
\pdfoutput=1

\usepackage{amsmath,amsthm,amssymb,euscript,epsf,epsfig}
\usepackage{array,latexsym,amsfonts,bm,enumerate,url}

\usepackage{graphicx}
\usepackage[usenames]{color}
\usepackage[english]{babel}

\usepackage{fancybox}

\usepackage{mciteplus}

\usepackage{parskip}     
\usepackage{titlesec}
\titlespacing{\section}{0pt}{3.5ex}{1ex}
\titlespacing{\subsection}{0pt}{2.5ex}{1ex}
\titlespacing{\subsubsection}{0pt}{2ex}{0ex}

\newcommand{\captionfonts}{\small}

\makeatletter  
\long\def\@makecaption#1#2{%
  \vskip\abovecaptionskip
  \sbox\@tempboxa{{\captionfonts #1: #2}}%
  \ifdim \wd\@tempboxa >\hsize
    {\captionfonts #1: #2\par}
  \else
    \hbox to\hsize{\hfil\box\@tempboxa\hfil}%
  \fi
  \vskip\belowcaptionskip}
\makeatother   

\usepackage[
      colorlinks=true,
      linkcolor=blue,
      urlcolor=blue,
      filecolor=blue,
      citecolor=blue,
      pdfstartview=FitH,
      pdftitle={},
      pdfauthor={},
      pdfsubject={},
      pdfkeywords={},
      pdfpagemode=UseNone,
      bookmarksopen=true
      ]{hyperref}
\usepackage[all]{hypcap}     

\begin{document}

\numberwithin{equation}{section}


\mathchardef\mhyphen="2D


\newcommand{\be}{\begin{equation}}
\newcommand{\ee}{\end{equation}}
\newcommand{\bea}{\begin{eqnarray}\displaystyle}
\newcommand{\eea}{\end{eqnarray}}
\newcommand{\nnm}{\nonumber}
\newcommand{\nn}{\nonumber}

\def\eq#1{(\ref{#1})}
\newcommand{\secn}[1]{Section~\ref{#1}}

\newcommand{\tbl}[1]{Table~\ref{#1}}
\newcommand{\fig}{Fig.~\ref}

\def\beq{\begin{equation}}
\def\eeq{\end{equation}}
\def\beqa{\begin{eqnarray}}
\def\eeqa{\end{eqnarray}}
\def\bet{\begin{tabular}}
\def\eet{\end{tabular}}
\def\bs{\begin{split}}
\def\es{\end{split}}


\def\a{\alpha}  
\def\b{\beta}  
\def\c{\chi}
\def\g{\gamma}
\def\G{\Gamma}
\def\e{\epsilon}
\def\vep{\varepsilon}
\def\tvep{\widetilde{\varepsilon}}
\def\f{\phi}
\def\F{\Phi}
\def\fb{{\ov \phi}}
\def\vf{\varphi}
\def\m{\mu}
\def\mub{\ov \mu}
\def\n{\nu}
\def\nub{\ov \nu}
\def\o{\omega}
\def\O{\Omega}
\def\r{\rho}
\def\k{\kappa}
\def\kab{\ov \kappa}
\def\s{\sigma}
\def\t{\tau}
\def\th{\theta}
\def\sb{\ov\sigma}
\def\S{\Sigma}
\def\l{\lambda}
\def\L{\Lambda}
\def\p{\psi}


\def\cA{{\cal A}} \def\cB{{\cal B}} \def\cC{{\cal C}}
\def\cD{{\cal D}} \def\cE{{\cal E}} \def\cF{{\cal F}}
\def\cG{{\cal G}} \def\cH{{\cal H}} \def\cI{{\cal I}}
\def\cJ{{\cal J}} \def\cK{{\cal K}} \def\cL{{\cal L}}
\def\cM{{\cal M}} \def\cN{{\cal N}} \def\cO{{\cal O}}
\def\cP{{\cal P}} \def\cQ{{\cal Q}} \def\cR{{\cal R}}
\def\cS{{\cal S}} \def\cT{{\cal T}} \def\cU{{\cal U}}
\def\cV{{\cal V}} \def\cW{{\cal W}} \def\cX{{\cal X}}
\def\cY{{\cal Y}} \def\cZ{{\cal Z}}

\def\mC{\mathbb{C}} 
\def\mP{\mathbb{P}}  
\def\mR{\mathbb{R}} 
\def\mZ{\mathbb{Z}} 
\def\mT{\mathbb{T}} 
\def\mN{\mathbb{N}}
\def\mH{\mathbb{H}}
\def\mX{\mathbb{X}}

\def\C{\mathbb{C}}
\def\CP{\mathbb{CP}}
\def\R{\mathbb{R}}
\def\RP{\mathbb{RP}}
\def\Z{\mathbb{Z}}
\def\N{\mathbb{N}}
\def\H{\mathbb{H}}

\newcommand{\rmd}{\mathrm{d}}
\newcommand{\rmx}{\mathrm{x}}

\def\tA{ {\widetilde A} } 

\def\one{{\hbox{\kern+.5mm 1\kern-.8mm l}}}
\def\zero{{\hbox{0\kern-1.5mm 0}}}


\newcommand{\red}[1]{{\color{red} #1}}
\newcommand{\blue}[1]{{\color{blue} #1}}
\newcommand{\green}[1]{{\color{green} #1}}

\definecolor{orange}{rgb}{1,0.5,0}
\newcommand{\orange}[1]{{\color{orange} #1}}


\newcommand{\bra}[1]{{\langle {#1} |\,}}
\newcommand{\ket}[1]{{\,| {#1} \rangle}}
\newcommand{\braket}[2]{\ensuremath{\langle #1 | #2 \rangle}}
\newcommand{\Braket}[2]{\ensuremath{\langle\, #1 \,|\, #2 \,\rangle}}
\newcommand{\norm}[1]{\ensuremath{\left\| #1 \right\|}}
\def\corr#1{\left\langle \, #1 \, \right\rangle}
\def\vac{|0\rangle}


\def\d{ \partial } 
\def\zb{{\bar z}}

\newcommand{\sq}{\square}
\newcommand{\IP}[2]{\ensuremath{\langle #1 , #2 \rangle}}    

\newcommand{\floor}[1]{\left\lfloor #1 \right\rfloor}
\newcommand{\ceil}[1]{\left\lceil #1 \right\rceil}

\newcommand{\dbyd}[1]{\ensuremath{ \frac{\d}{\d {#1}}}}
\newcommand{\ddbyd}[1]{\ensuremath{ \frac{\d^2}{\d {#1}^2}}}

\newcommand{\Zd}{\ensuremath{ Z^{\dagger}}}
\newcommand{\Xd}{\ensuremath{ X^{\dagger}}}
\newcommand{\Ad}{\ensuremath{ A^{\dagger}}}
\newcommand{\Bd}{\ensuremath{ B^{\dagger}}}
\newcommand{\Ud}{\ensuremath{ U^{\dagger}}}
\newcommand{\Td}{\ensuremath{ T^{\dagger}}}

\newcommand{\T}[3]{\ensuremath{ #1{}^{#2}_{\phantom{#2} \! #3}}}		

\newcommand{\tr}{\operatorname{tr}}
\newcommand{\Str}{\operatorname{Str}}
\newcommand{\sech}{\operatorname{sech}}
\newcommand{\Spin}{\operatorname{Spin}}
\def\Tr{{\rm Tr\, }} 
\newcommand{\Sym}{\operatorname{Sym}}
\newcommand{\Com}{\operatorname{Com}}
\def\adj{\textrm{adj}}
\def\id{\textrm{id}}
\def\Id{\textrm{Id}}
\def\ind{\textrm{ind}}
\def\Dim{\textrm{Dim}}
\def\End{\textrm{End}}
\def\Res{\textrm{Res}}
\def\Ind{\textrm{Ind}}
\def\ker{\textrm{ker}}
\def\im{\textrm{im}}
\def\sgn{\textrm{sgn}}
\def\ch{\textrm{ch}}
\def\STr{\textrm{STr}}
\def\Sym{\textrm{Sym}}

\def\ha{\frac{1}{2}}
\def\tha{\tfrac{1}{2}}
\def\wt{\widetilde}
\def\ra{\rangle}
\def\la{\langle}

\def\pb{\ov\psi}
\def\pt{\widetilde{\psi}}
\def\at{\widetilde{\a}}
\def\cb{\ov\chi}
\def\d{\partial}
\def\db{\bar\partial}
\def\delb{\bar\partial}
\def\dbar{\ov\partial}
\def\dag{\dagger}
\def\dalpha{{\dot\alpha}}
\def\dbeta{{\dot\beta}}
\def\dgamma{{\dot\gamma}}
\def\ddelta{{\dot\delta}}
\def\ad{{\dot\alpha}}
\def\bd{{\dot\beta}}
\def\dg{{\dot\gamma}}
\def\dd{{\dot\delta}}
\def\th{\theta}
\def\Th{\Theta}
\def\eb{{\ov \epsilon}}
\def\gb{{\ov \gamma}}
\def\wb{{\ov w}}
\def\Wb{{\ov W}}
\def\ib{{\ov i}}
\def\jb{{\ov j}}
\def\kb{{\ov k}}
\def\mb{{\ov m}}
\def\nb{{\ov n}}
\def\qb{{\ov q}}
\def\Qb{{\ov Q}}
\def\xh{\hat{x}}
\def\D{\Delta}
\def\DD{\Delta^\dag}
\def\Db{\ov D}
\def\M{{\cal M}}
\def\rd{\sqrt{2}}
\def\ov{\overline}
\def\Slash{\, / \! \! \! \!}
\def\dslash{\partial\!\!\!/} 
\def\Dslash{D\!\!\!\!/\,\,}
\def\fslash#1{\slash\!\!\!#1}
\def\Fslash#1{\slash\!\!\!\!#1}

\def\del{\partial}
\def\delb{\bar\partial}
\newcommand{\ex}[1]{{\rm e}^{#1}} 
\def\ii{{i}}

\renewcommand{\theequation}{\thesection.\arabic{equation}}
\newcommand{\vs}[1]{\vspace{#1 mm}}

\newcommand{\ve}{{\vec{\e}}}
\newcommand{\shalf}{\frac{1}{2}}

\newcommand{\lb}{\rangle}
\newcommand{\al}{\ensuremath{\alpha'}}
\newcommand{\ap}{\ensuremath{\alpha'}}

\newcommand{\bean}{\begin{eqnarray*}}
\newcommand{\eean}{\end{eqnarray*}}
\newcommand{\ft}[2]{{\textstyle {\frac{#1}{#2}} }}

\newcommand{\hsp}{\hspace{0.5cm}}
\def\half{{\textstyle{1\over2}}}
\let\ci=\cite \let\re=\ref
\let\se=\section \let\sse=\subsection \let\ssse=\subsubsection

\newcommand{\dpb}{D$p$-brane}
\newcommand{\dpbs}{D$p$-branes}

\def\gh{{\rm gh}}
\def\sgh{{\rm sgh}}
\def\NS{{\rm NS}}
\def\R{{\rm R}}
\def\Qp{Q_{\rm P}}
\def\QP{Q_{\rm P}}

\newcommand\dott[2]{#1 \! \cdot \! #2}

\def\eo{\overline{e}}



\def\p{\partial}
\def\h{{1\over 2}}

\def\d{\partial}
\def\la{\lambda}
\def\eps{\epsilon}
\def\bb{\bigskip}
\def\mm{\medskip}
\def\tg{\widetilde\gamma}
\newcommand{\dm}{\begin{displaymath}}
\newcommand{\edm}{\end{displaymath}}
\renewcommand{\b}{\widetilde{B}}
\newcommand{\gm}{\Gamma}
\newcommand{\ac}[2]{\ensuremath{\{ #1, #2 \}}}
\renewcommand{\ell}{l}
\newcommand{\z}{\ell}
\newcommand{\newsection}[1]{\section{#1} \setcounter{equation}{0}}
\def\bb{$\bullet$}
\def\Qbar{{\bar Q}_1}
\def\QPbar{{\bar Q}_p}

\def\q{\quad}

\def\bn{B_\circ}

\let\vp=\varphi \let\vep=\varepsilon
\let\w=\omega      
\let\G=\Gamma \let\D=\Delta \let\Th=\Theta
\let\P=\Pi \let\S=\Sigma

\def\h{{1\over 2}}
\def\r{\rightarrow}
\def\nn{\nonumber\\}
\let\bm=\bibitem
\def\Kt{{\widetilde K}}
\def\b{\bigskip}

\let\p=\partial


%
%
\begin{flushright}
\end{flushright}

\vspace{1.5cm}

\centerline{
\LARGE{ {\bf Black hole microstate geometries} } }
\vspace{0.5cm}

\centerline{
\LARGE{ {\bf from string amplitudes\footnote{(Expanded version of) talk given at the Black Objects in Supergravity School, Frascati, 2011}}}}

\vspace{2cm}

\centerline{    
{\bf David Turton} }

\vspace{0.4cm}

\begin{center}
Department of Physics,\\ The Ohio State University,\\ Columbus,
OH 43210, USA

\vspace{0.4cm}

turton.7@osu.edu

\end{center}

\vspace{2cm}

\centerline{
{\bf Abstract} }
 
\vspace{4mm}

In this talk we review recent calculations of the asymptotic supergravity fields sourced by bound states of D1 and D5-branes carrying travelling waves.
We compute disk one-point functions for the massless closed string fields.
At large distances from the branes, the effective open string coupling is small, even in the regime of parameters where the classical D1-D5-P black hole may be considered.
The fields sourced by the branes differ from the black hole solution by various multipole moments, and have led to the construction of a new $1/8$-BPS ansatz in type IIB supergravity.

\thispagestyle{empty}

\newpage

\section{Introduction}

Black holes provide (at least) two major challenges for any theory of quantum gravity: to give a microscopic interpretation of the Bekenstein-Hawking entropy~\cite{Bekenstein:1973ur,Hawking:1974sw}, and to resolve the information paradox~\cite{Hawking:1976ra}.
String theory promises to pass both tests: the microscopic interpretation of the Bekenstein-Hawking entropy is provided by enumerating microstates of the black hole~\cite{Sen:1995in,Strominger:1996sh,Callan:1996dv}, and studying the properties of these microstates promises to resolve the information paradox.

The information paradox states, roughly, that if a classical black hole metric with a horizon is a valid description of a physical black hole in Nature, then Hawking radiation leads to a breakdown of unitarity or exotic remnant objects (for a recent rigorous treatment, see~\cite{Mathur:2009hf}). A conservative way of avoiding these pathologies is to ask whether the physics of individual black hole microstates modifies the process of Hawking radiation.

The study of the gravitational description of individual microstates has motivated a `fuzzball' picture of a black hole~\cite{Mathur:2005zp,Mathur:2008nj}.
The fuzzball conjecture is composed of two parts: firstly that quantum effects are important at the would-be horizon of a black hole, making Hawking radiation a unitary process;
secondly, that the mechanism underlying this is that the quantum bound state of matter making up the black hole has a macroscopic size, of order the horizon scale.

To investigate this conjecture, one must understand the characteristic size of (the wavefunctions of) individual bound states. A fruitful line of inquiry has been to construct and analyze classical supergravity solutions describing the gravitational fields sourced by semiclassical/coherent states of the Hilbert space of the black hole (for reviews, see~\cite{Bena:2007kg,Skenderis:2008qn,Balasubramanian:2008da,Chowdhury:2010ct}). Supergravity solutions which describe individual microstates have been found not to have horizons themselves. 

Given such a supergravity solution however, it may not always be clear whether it corresponds to a black hole microstate (see e.g.~\cite{Bena:2006kb,Sen:2009bm}).
In this talk we describe calculations which directly associate supergravity fields with the microscopic bound states they describe. We consider particular bound states of D-branes, and derive the asymptotic supergravity fields from worldsheet amplitudes. The amplitudes are disk-level one-point functions for the emission of massless closed string fields.

We first derive the fields sourced by a D1-brane with a travelling wave and relate them to the previously known two-charge supergravity fields~\cite{Dabholkar:1995nc,Callan:1995hn}. We then derive the fields sourced by a D1-D5 bound state with a travelling wave and find a new set of three-charge supergravity fields, more general than previously considered~\cite{Bena:2004de}.
The results reviewed here appeared in the papers~\cite{Black:2010uq} and~\cite{Giusto:2011fy}.

This talk is structured as follows. In Section \ref{sec:validity} we introduce the calculation and discuss its regime of validity. In Section \ref{sec:D1-P} we review the D1-P calculation, and in Section \ref{sec:D1-D5-P} we review the D1-D5-P calculation.

\section{The calculation and its regime of validity} \label{sec:validity}

The procedure we follow for calculating the asymptotic fields sourced by D-brane bound states was developed in~\cite{garousi96,hashimoto96,DiVecchia:1997pr,Giusto:2009qq}. First, we calculate the momentum-space amplitude $\cA(k)$ for the emission of a massless closed string. We then extract the field of interest (e.g.~graviton), multiply by a free propagator, and Fourier transform to obtain the spacetime one-point function.

\begin{figure}[h]
\begin{center}
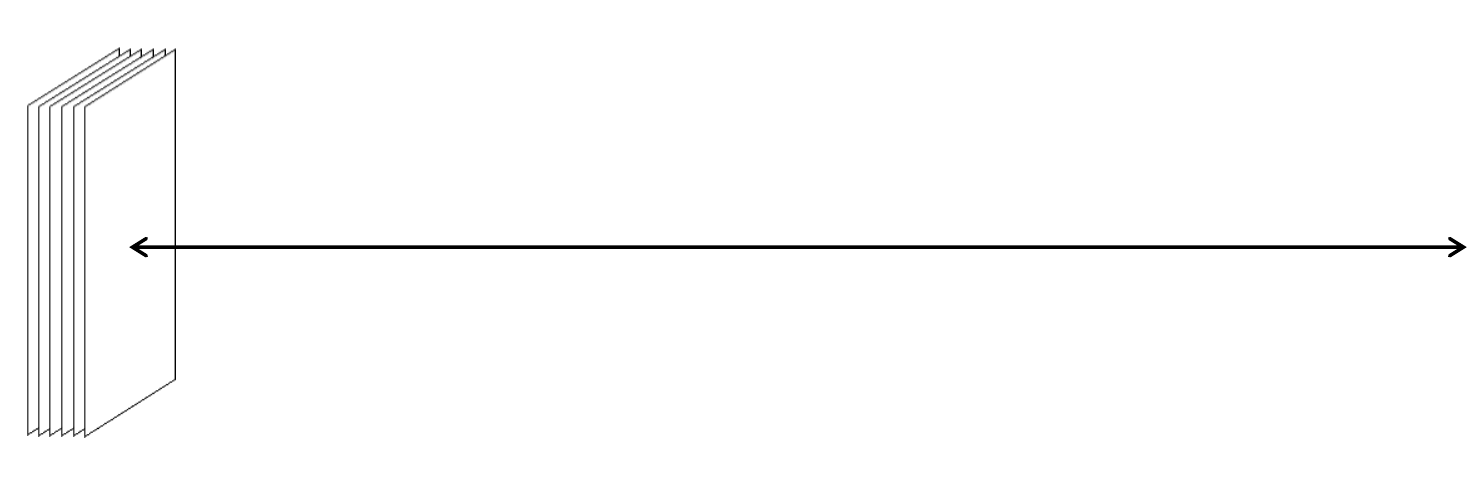
\end{center}
\caption{\label{Large_dist} For the calculation of the fields at large distances $r$ from a bound state of $N$ D$p$-branes, the effective open string coupling is small if $r^{7-p} \gg g_s N \sqrt{\ap}^{7-p}$.}
\end{figure}

For applications to black holes, given $N$ D-branes we are interested in the regime $g_s N \gg 1$, where a classical black hole solution might be relevant. The naive open string coupling is also $g_s N$, so it seems we are out of the regime of open string perturbation theory (see e.g.~\cite{Strominger:1996sh}).

However if one considers the above calculation for the fields at a distance $r$ from the bound state (Fig.\,\ref{Large_dist}), one finds that the effective open string coupling is in fact 
\be \label{eq:epsilon}
\e ~=~ 
g_s N \left( \frac{\ap}{r^2} \right)^{\!\frac{7-p}{2}} \,.
\ee
This effective open string coupling may be understood as follows.
The next order in open string perturbation theory corresponds to adding an extra border to the string worldsheet. 
The factor of $N$ comes from the $N$ choices of which D$p$-brane the open string endpoints can end on. The extra border on the worldsheet also introduces a loop momentum integral, two extra propagators, and reduces the background superghost charge by two units. This results in the above powers of $\frac{\ap}{r^2}$, as discussed in detail in~\cite{Giusto:2011fy}. 

The next order in closed string perturbation theory corresponds to adding handles to the closed string propagator, which we suppress by working at $g_s \ll 1$. Thus we work in the following regime of parameters: 
\be
g_s \, \ll \, 1~, \qquad 
g_s N \left( \frac{\ap}{r^2} \right)^{\!\frac{7-p}{2}}  \ll ~ 1 ~.
\ee
Thus one can simultaneously consider $g_s N \gg 1$, provided $r$ is sufficiently large. 

One can rephrase the second condition above as saying that disk amplitudes give the leading contribution to the fields at 
lengthscales much greater than the characteristic size of the D-brane bound state, $r^{7-p} \gg g_s N \sqrt{\ap}^{7-p}$.
A similar perturbative expansion was made some time ago in the field theory analogue of our calculation~\cite{Duff:1973zz}.

Since the fields in which we are interested are massless, the emitted closed string state has non-zero momentum only in the four non-compact directions of the $\mathbb{R}^4$, i.e. a spacelike momentum. The momentum-space amplitude $\cA(k)$ mentioned above is defined by analytically continuing $k$ to complex values such that we impose $k^2=0$, i.e. the emitted string state is treated as on-shell~\cite{DiVecchia:1997pr}.

One can ask whether this procedure fails to capture any physics relevant to the calculation. For example, one could add to the amplitude $\cA(k)$ a contribution proportional to any positive power of $k^2$, which would vanish if $k^2=0$. Suppose we add a term proportional to $k^2$; then multiplying by a free propagator $1/k^2$ and Fourier transforming gives a Dirac delta-function in position space. Similarly, higher powers of $k^2$ correspond to derivatives of the delta-function in position space. This signifies that these terms are relevant for physics very close to the location of the D-brane, and do not affect the large distance behaviour of the supergravity fields.


\section{The two-charge D1-P amplitude} \label{sec:D1-P}

We consider type IIB string theory on $\mathbb{R}^{1,4}\times
S^1\times T^4$. We denote the 10D coordinates $(x^{\m},\psi^{\m})$
by $ \m, \n =t,y,1,\ldots,8$. We use $(i,j,\ldots)$ and $x^1, \ldots , x^4$ for the $\mathbb{R}^{4}$ directions, we use $(a,b,\ldots) \,$ and $x^5, \ldots , x^8$ for the $T^4$ directions
and we use $(I,J,\ldots)$ to refer to the combined $\mathbb{R}^{1,4}\times S^1$ directions.
We work in the light-cone coordinates
\be
v = \left( t + y \right), \qquad  u = \left( t - y \right) ~
\ee
constructed from the time and $S^1$ directions. We consider a D1-brane wrapped around $y$ and carrying a $v$-dependent travelling wave:
\be
\begin{array}{c|cc|c|c}
   &   v  &   u  & \mR^4  &  T^4   \\ 
D1 & \rmx & \rmx & f_i(v) & f_a(v)=0  \\
\end{array}
\ee
Here ``x'' denotes a Neumann direction and $f$ indicates the
($v$-dependent) position of the D-brane in the Dirichlet directions.
From the start we set the profile along the $T^4$ directions to be trivial, $ f_a = 0$. An analogous calculation may be performed for the case of a D5-brane wrapped on the $T^4 \times S^1$ directions, and both these amplitudes contribute to the D1-D5-P amplitude that we discuss in the next section.

\begin{figure}[h]
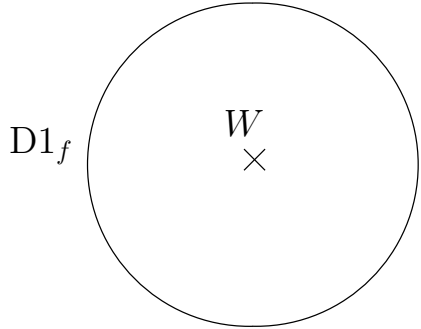

\begin{center}
\input twoc.pdf_t
\end{center}
\caption{\label{Dpfig} The one-point function for emission of the closed string state $W$ from a disk ending on a D1 brane with profile $f$.}
\end{figure}

The boundary conditions on the worldsheet fields in the open string picture may be expressed in terms of a reflection matrix $R$ as 
\bea
\pt^\m &=& \eta \, \T{R}{\m}{\n}(\mathbf{v}) \psi^\n    \label{psii} \\
\dbar X_R^\m &=&  \T{R}{\m}{\n}(\mathbf{v}) \d X_L^\n - \T{\delta}{\m}{u} \, 4
\ap \ddot{f}_j(\mathbf{v}) \psi^j \psi^v 
\label{nln}
\eea
where $\eta$ can be set to 1 at $\s=0$, while at $\s=\pi$ we have $\eta = 1$ or $\eta = -1$ corresponding to
the NS and R sectors respectively. In the above we use a bold letter for the string field corresponding to the coordinates 
\be
\mathbf{x}^{\mu}(z,\zb) ~=~ \frac{1}{2} \Big[ X_L(z) + X_R(\zb) \Big] \,.
\ee
The holomorphic and the anti-holomorphic world-sheet
fields are then identified with the reflection matrix $R$
where (see~\cite{Black:2010uq} and references within)
\be \label{eq:D1_bcs}
R^{\m}_{\phantom{\m \!}\n} ~=~ 
\left( \begin{array}{cccc} 
        	  1              &      0      &          0           &          0          \\
   	  4 |\dot{f}({\mathbf{v}})|^2   &      1      &  -4 \dot{f}_i({\mathbf{v}})   &          0          \\
 	    2 \dot{f}_i({\mathbf{v}})   &      0      &    \!\! - \one       &          0          \\
  	        0              &      0      &          0           &   \!\! - \one
\end{array} \right) \;,
\ee 
where $\one$ denotes the four-dimensional unit matrix and the indices follow the ordering $(v,u,i,a)$.

The most direct way to derive the one-point functions in the current setup is to use the boundary state
formalism~\cite{DiVecchia:1997pr}.  The calculation we now review was carried out in~\cite{Black:2010uq} by using the boundary state for a D-brane with a null wave derived in~\cite{Hikida:2003bq,*Blum:2003if,Bachas:2003sj}. 

The wrapped D1-brane may be viewed as as a set of $n_w$ different D-brane strands, with a non-trivial holonomy gluing these strands together. Each strand carries a segment of the full profile $f^i_{(s)}$, with $s=1,\ldots,n_w$. The boundary state describing the wrapped D1-brane can be expanded in terms of the closed string perturbative states. The first terms of this expansion are
\begin{eqnarray} 
\label{eq:BdyStateStrands}
\ket{D1;f} & = &  -i \frac{ \k \, \t_{1} }{2} 
\sum_{s=1}^{n_w} \int \! du \int\limits_0^{2 \pi R} \! d{v} \int \!
\frac{d^4p_i}{(2 \pi)^4} \,e^{-i p_i f^i_{(s)}({v}) } 
\frac{c_0+\widetilde{c}_0}{2} 
\\ &  &\nonumber \times ~c_1 \widetilde{c}_1  \left[
  -\psi^\mu_{-\frac 12} ({}^{\rm t}\! R)_{\mu\nu}
  \widetilde\psi^\nu_{-\frac 12} + \gamma_{-\frac 12} \widetilde\beta_{-\frac
    12} -  \beta_{-\frac 12} \widetilde\gamma_{-\frac 12} +\ldots
\right] \ket{u,v,p_i,0}_{-1,\widetilde{-1}}\, \quad
\end{eqnarray}
where $\t_{1}= [2\pi \ap g_s]^{-1}$ is the physical tension of a D1-brane, and where ${}^{\rm t}\!R$ is the transpose of $R$. The ket in~\eq{eq:BdyStateStrands} represents a closed string state obtained by acting on the $SL(2,C)$ invariant vacuum with an $e^{i p_i   x^i}$ in the $\mathbb{R}^{4}$ directions. We wrote the delta functions on the $p_u$ and $p_v$ momenta as integrals in configuration space $du,\,dv$. The boundary state enforces the identification~\eqref{psii}, which in the
approximation~\eqref{eq:BdyStateStrands} holds just for the first oscillator $\widetilde\psi^\mu_{-1/2}$.

The second line of~\eqref{eq:BdyStateStrands} contains all the massless NS-NS states; we can separate the irreducible
contributions by taking the scalar product with each state. Having done this, the contribution to the
NS-NS couplings is
\begin{eqnarray}
 \cA_{\rm dil}(k) &=& 
 - i \frac{\k \, \t_{1} }{2} V_u
\sqrt{2}\hat{\phi} \int\limits_0^{\,\,L_T}
 \! d\hat{v}  \,  e^{- i k \cdot f (\hat{v}) } ~,\\
\cA_{\rm gra}(k)& = &
 - i \frac{\k \, \t_{1} }{2} V_u \int\limits_0^{\,\,L_T}
 \! d\hat{v}  \,  e^{- i k \cdot f  (\hat{v}) }  
\left[-\frac{3}{2} (-\hat{h}_{tt} +\hat{h}_{yy}) \right.
\\ \nonumber & & \qquad\qquad  \left. + \frac{1}{2}
   (\hat{h}_{ii} + \hat{h}_{aa}) -2 \hat{h}_{vv} |\dot{f}|^2 +
   4\hat{h}_{vi} \dot{f}^i\right] ~
\end{eqnarray}
where $V_u$ is the (divergent) volume along the $u$ direction and the integrals over $v$ in each strand in~\eqref{eq:BdyStateStrands} have become a single integral over the multi-wound worldvolume coordinate $\hat{v}$ which runs from $0$ to
$L_T=2\pi n_w R$.

For the R-R coupling, we simply recall the results of~\cite{Black:2010uq}:
\begin{eqnarray}
  \label{eq:RRD1f}
  \cA_{\rm RR}(k) & = &
  - i \sqrt{2} \k \, \t_{1} 
  V_u \int\limits_0^{\,\,L_T}
  \! d\hat{v}  \,  e^{- i k \cdot f  (\hat{v}) }  
  \left[2 \hat{C}^{(2)}_{uv} + \hat{C}^{(2)}_{vi} \dot{f}^i\right].
\end{eqnarray}

The next step is to multiply by a free propagator and Fourier transform to find the position-space massless fields. After doing this, one finds agreement with the known D1-P solutions obtained by an S-duality of the solutions of~\cite{Dabholkar:1995nc,Callan:1995hn}. Further details may be found in~\cite{Black:2010uq}.


\section{The three-charge D1-D5-P amplitude} \label{sec:D1-D5-P}

We next consider a D1-D5 bound state carrying a travelling wave; the black hole solution with the same charges has a macroscopic horizon~\cite{Strominger:1996sh}, and so this case is more interesting and richer than that of the previous section.

We consider a D1-D5 bound state with a common travelling-wave profile $f_i(v)$ along the branes. The D5-brane is wrapped on the $T^4 \times S^1$.
\be
\begin{array}{c|cc|c|c}
   &   v  &   u  & \mR^4  &  T^4   \\ 
D1 & \rmx & \rmx & f_i(v) & f_a(v)=0  \\
D5 & \rmx & \rmx & f_i(v)  & \rmx  \\
\end{array}
\ee
The disk amplitude of most interest in this setup is the one where the disk has half its boundary on a D1 and the other half on the D5, and two twisted vertex operator insertions, as studied in~\cite{Billo:2002hm,Giusto:2009qq}.

\begin{figure}[htp]
\begin{center}
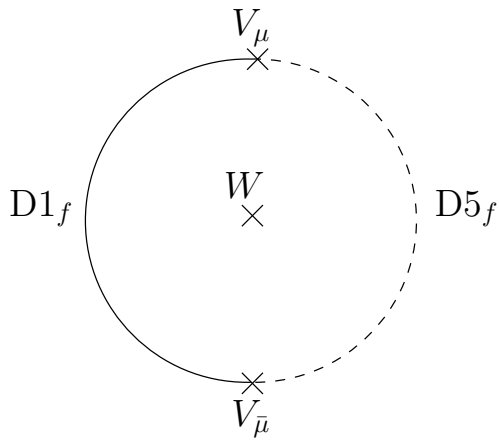
\end{center}
\caption{\label{Dmixfig} The simplest amplitude involving all three
  charges of the D1-D5-P microstate: the worldsheet topology is that of a mixed disk diagram where part of the
  border lies on the D1 brane and part on the D5 brane.}
\end{figure}

The vertex operators take the form
\beq\label{Vop15} 
V_{\mu} = \mu^{A} \ex{-\frac{\varphi}{2}} S_A \,
 \Delta \,, \quad\qquad ~~~~~~~
V_{\bar \mu} =\bar \mu^{A} \ex{-\frac{\varphi}{2}}  S_A \, \Delta \,
\eeq
where $\mu^A$ and $\bar{\mu}^A$ are Chan-Paton matrices with
$n_1\times n_5$ and $n_5\times n_1$ components respectively, $S_A$ are
the $SO(1,5)$ spin fields, $\varphi$ is the free boson appearing in the
bosonized language of the worldsheet superghost $(\beta,\gamma)$, and
$\Delta$ is the bosonic twist operator with conformal dimension
$\frac{1}{4}$ which acts along the four mixed ND directions
and changes the boundary conditions from Neumann to Dirichlet and vice versa.

We focus on open string condensates involving the Ramond sector states only. These states break the $SO(4)$ symmetry of the DD directions $\mathbb{R}^4$, and are invariant under the $SO(4)$ acting on the compact $T^4$ torus.
The most general condensate of Ramond open strings can be written as:
\be\label{mubarmu}
\bar \mu^A \, \mu^B=v_I (C\Gamma^I)^{[AB]}+
\frac{1}{3!} \,v_{IJK} (C\Gamma^{IJK})^{(AB)}~,
\ee
where the parentheses indicate that the first term is automatically antisymmetric, while the second is
symmetric. The open string bispinor condensate is thus specified by a one-form $v_I$ and an self-dual three-form $v_{IJK}$. The self-duality of $v_{IJK}$ follows from $\bar\mu^A$ and $\mu^B$ having definite 6D chirality and can be written as
\beq\label{sdv}
v_{IJK} = \frac{1}{3!} \epsilon_{IJKLMN} v^{LMN}\,.
\eeq
In this talk we consider only the components of $v_{IJK}$ which have one leg in the $t,y$ directions and two legs in the $\mR^4$; this choice of components was associated to considering profiles only in the $\mR^4$ directions in~\cite{Giusto:2009qq}.
Since the spinors $\bar \mu^A$ and $\mu^B$ carry $n_5\times n_1$ and $n_1\times n_5$ Chan-Paton indices, the condensate $\bar\mu^{A}\mu^{B}$ must be thought of as the vev for the sum
\be
  \sum_{m=1}^{n_1} \sum_{n=1}^{n_5}  \bar\mu^{A}_{m n}\,  \mu^{B}_{n m}\,,
\ee
which, for generic choices of the Chan-Paton factors, is of order $n_1 n_5$.

The open string insertions~\eqref{Vop15} are related to the vevs of the strings stretched between the D1 and D5 branes, which we are treating perturbatively.
The microstates for which we might expect a gravitational
description have large open string vevs, so in principle we
should resum amplitudes with many twisted vertices.  
However each pair of open string
insertions~\eqref{Vop15} comes with a factor of $1/r$ in the large
distance expansion of the corresponding gravity solution~\cite{Giusto:2009qq,Giusto:2011fy}.

Thus in the following we focus only on the leading contributions
at large distances which are induced by the amplitudes with one border and one pair of
open vertices $V_\mu$, $V_{\bar\mu}$ (see Figure~\ref{Dmixfig}). This should be
sufficient to derive the sourced supergravity fields up to order $1/r^4$.

Thus the amplitude we now calculate is
\begin{equation}\label{c}
  {\cal A}_{NS,R}^{\rm D1\mhyphen D5} = \int \frac{\prod_{i=1}^4
    dz_i}{dV_{\mathrm{CKG}}} \, \left\langle V_\mu(z_1) \,
    W_{NS,R}^{(-k)}(z_2, z_3) \,
    V_{\bar \mu} (z_4) \,
  \right\rangle_f~,
\end{equation} 
where the subscript $f$ is to remind that, in this disk correlator, the
identification between holomorphic and anti-holomorphic components
depends on the profile of the D-branes through \eq{psii}--\eq{eq:D1_bcs}.

In order to have a non-trivial correlator we must saturate the superghost charge ($-2$) of
the disk. The two open string vertices together contribute $-1$, thus in the NS sector we use the closed string vertex operator in the $(0,-1)$ picture,
\begin{equation}
W_{NS}^{(k)} =  {\cal G}_{\mu\nu}
 \left(\partial X^{\mu}_L -
\ii \;\! \frac{k}{2} \! \cdot \! \psi \, \psi^{\mu} \right)
 \,\ex{ \ii \frac{k}{2} \cdot X_L} (z)
\, \widetilde\psi^{\nu} 
\ex{-\widetilde\varphi} \ex{ \ii  \frac{k}{2} \cdot X_R }(\bar z) + \ldots \,,
\label{clNS0}
\end{equation}
where the dots stand for other terms that ensure the BRST invariance
of the vertex, but that do not play any role in the correlator under
analysis.

We will not review the intermediate steps of the calculation here; details are given in~\cite{Giusto:2011fy}. We move on to discuss the spacetime fields which result from the calculation.

\subsection{New D1-D5-P geometries}

The fields obtained from the calculation of the previous section fit into the following ansatz, which solves the supergravity equations perturbatively\footnote{This ansatz was later extended to a full non-linear supergravity ansatz in~\cite{Giusto:2012gt}. Solutions have been studied in~\cite{Vasilakis:2012zg}.} in $1/r$ up to $1/r^4$.
Using the short-hand notation
\be
d\hat t = dt+k \,,\quad d\hat y = dy+dt-\frac{dt+k}{Z_3}+a_3\,,\label{dtdy}
\ee
the ansatz (in the string frame) is
\bea
ds^2 &=& \frac{1}{\sqrt{Z_1 Z_2}}\Bigl[-\frac{1}{Z_3}\, d{\hat t}^2 + Z_3\, d{\hat y}^2 \Bigr]+\sqrt{Z_1 Z_2}\, ds^2_4+\sqrt{\frac{Z_1}{Z_2}}\, ds^2_{T^4} \,, \cr
B &=& -Z_4\,d\hat t \wedge d\hat y + a_4\wedge ( d \hat t + d\hat y) + \delta_2\,, \cr
e^{2\phi} &=& \frac{Z_1}{Z_2}\,, \label{NSNSsusy} \qquad 
C^{(0)} ~=~ Z_4 \,, \cr
C^{(2)} &=& -\frac{1}{Z_1} d\hat t \wedge d\hat y + a_1\wedge ( d \hat t + d\hat y) + \gamma_2\,,\label{RRsusy} \\
F^{(5)} &=& dZ_4 \wedge dz^4  +\frac{Z_2}{Z_1}\, *_4 dZ_4\wedge d\hat t \wedge d\hat y\,, \nnm
\eea
where $ds^2_4$ is a generic Euclidean metric on $\mathbb{R}^4$; $ds^2_{T^4}$ is the flat metric on $T^4$; $Z_I$ are 0-forms, $k, a_I$ are 1-forms, and $\gamma_2, \delta_2$ are 2-forms on $\mathbb{R}^4$. The above quantities are subject to the conditions
\be
d \delta_2 = *_4 d a_4\,,\quad d\gamma_2 = *_4  d Z_2\,,\label{deltagamma}
\ee
and we take the asymptotic boundary conditions
\bea\label{asymptoticbc}
Z_1,\, Z_2, \,Z_3 &=& 1+ O(r^{-2})\,, \qquad  Z_4 ~=~ O(r^{-4}) \,,\cr
k,\, a_1,\, a_3,\, a_4 &=& O(r^{-3}) \,,  \qquad\quad~ ds^2_4 ~=~ dx_i dx_i + O(r^{-4}) \,.
\eea

The above fields satisfy the approximate supergravity Killing spinor equations up to order $1/r^4$.  
It turns out, however, that one can keep the full $r$ dependence of the string results and still
satisfy the approximate supergravity Killing spinor equations, to linear order in the condensate $v_{IJK}$.

The full $r$ dependence of the supergravity fields describes
the small $g_s N$ and small $\mathbf{v}_{IJK}$ limit, i.e. the weak gravity regime and 
the region of the Higgs branch infinitesimally close to its intersection with the Coulomb branch. If one 
is interested in the black hole regime (large $g_s N$ and finite $\mathbf{v}_{IJK}$), 
one should keep only the large $r$ limit (up to $1/r^4$ order) of the following results. 

We next review the most interesting features of the string amplitude; for the full set of fields see~\cite{Giusto:2011fy}.
If we set $Z_4$ and $a_4$ to zero, the ansatz reduces to that in~\cite{Bena:2004de}. These `new' fields are thus the most interesting. The disk amplitude gives 
\bea
Z_4 &=& - \mathbf{v}_{ujk} \, \p_j \left[ \frac{1}{L_T} \int_0^{L_T} \!\!d\hat v\,\frac{\dot{f}_k}{|x^i-f^i|^2} \right] \,, \\ 
a_4 &=&  \mathbf{v}_{uij} \, \p_j \left[ \frac{1}{L_T} \int_0^{L_T} \!\!d\hat v\,\frac{|\dot{f}|^2}{|x^i-f^i|^2} \right] dx^i
\eea
where we have absorbed some factors multiplying the open string condensate,
\begin{equation}
  \mathbf{v}_{IJK} ~=~ -\frac{2 \sqrt{2} n_w \kappa}{\pi V_4} v_{IJK} \,.
\end{equation}
Note that the new fields above vanish in either of the two-charge limits in which we set either $v_{IJK}$ or $f$ to zero.

Another interesting outcome of our calculation is that it predicts that the 4D base metric $ds^2_4$, which is simply the flat metric on $\mathbb{R}^4$ in the 2-charge case, is a {\em non-trivial} hyper-K\"ahler metric when all three charges are present. The base metric which arises from the string amplitude is
\bea
ds^2_4 &=& \Big( \delta_{ij} + \mathbf{v}_{uli}\,\partial_l \mathcal{I}_j +\mathbf{v}_{ulj}\,\partial_l \mathcal{I}_i - \delta_{ij} \,\mathbf{v}_{ulk}\,\partial_l \mathcal{I}_k \Big) \,dx^i dx^j\,,\label{basemetric}
\eea
where
\bea
\mathcal{I}_j &=& \frac{1}{L_T} \int_0^{L_T} \!\!d\hat v\,\frac{\dot{f}_j}{|x^i-f^i|^2}\,.
\eea
The non-flatness of the base metric for 3-charge microstate geometries was previously observed in the particular solution of \cite{Giusto:2004kj}, but had remained until now largely unexplained. It is nice to see that 
the disk amplitudes lead directly to this feature.

\newpage 

\section{Summary} \label{sec:Discussion}

In this talk we have seen how disk amplitudes can be used to derive the asymptotic supergravity fields sourced by bound states of D-branes. At large distances from the bound state, the effective open string coupling is small,
even in the regime of parameters in which there is a classical black hole solution with the same charges. 

The supergravity fields differ from the black hole solution by various multipole moments, suggesting that the D1-D5-P black hole solution is not an exact description of the gravitational fields sourced by individual microstates. Rather the black hole solution is likely to be an approximate thermodynamic description of the entire system. Thus the results reviewed here support the fuzzball proposal.

It would be interesting to apply the techniques reviewed here to other D-brane bound states, and 
we hope that this will lead to an improved understanding of the physics of black holes in string theory.


\section*{Acknowledgements}

I thank W.~Black, S.~Giusto, and R.~Russo for collaboration on the research reviewed here and
I thank V.~Jejjala, S.~D.~Mathur, J.~F.~Morales, S.~Ramgoolam, A.~Sen and E.~Witten for fruitful discussions. 
I would like to acknowledge the support of an STFC studentship at Queen Mary, University of London and of DOE grant DE-FG02-91ER-40690.


\providecommand{\href}[2]{#2}\begingroup\raggedright\endgroup

\end{document}